\UseRawInputEncoding
\documentclass[pra,twocolumn]{revtex4}
\usepackage{amsmath}
\usepackage{amssymb}
\usepackage{dsfont}
\usepackage{color}
\makeatletter
\usepackage{amssymb}
\usepackage{graphicx}
\usepackage{epstopdf}
\usepackage{epsfig}
\usepackage{subfigure}
\usepackage[colorlinks,urlcolor=blue,citecolor=blue]{hyperref}
\makeatother

\newcommand{\ket}[1]{|#1\rangle}

\newcommand{\e}{\text{e}}

\usepackage{cancel,ifthen}
\usepackage[normalem]{ulem}

\newcommand{\me}[1]{{\color{black} #1}}
\newcommand{\iref}[1]{{\it #1}}

\newcommand{\comment}[2][NoInPuT]{\ifthenelse{\equal{#1}{NoInPuT}}{}{{\color{blue}\sout{#1}}}{\color{red} #2}}
\begin{document}

\title{Few-to-many vortex states of density-angular-momentum coupled Bose-Einstein condensates}

\author{Matthew Edmonds}
\affiliation{Department of Physics \& Research and Education Center for Natural Sciences, Keio University, Hiyoshi 4-1-1, Yokohama, Kanagawa 223-8521, Japan}

\begin{abstract}
\noindent Motivated by recent experiments, we theoretically study a gas of atomic bosons confined in an elliptical harmonic trap; forming a quasi-two-dimensional atomic Bose-Einstein condensate subject to a density-dependent gauge potential which realises an effective density-angular-momentum coupling. We present exact Thomas-Fermi solutions which allows us to identify the stable regimes of the full parameter space of the model. Accompanying numerical simulations reveal the effect of the interplay of the rigid body and density-angular-momentum coupling for the elliptically confined condensate. By varying the strength of the gauge potential and trap anisotropy we explore how the superfluid state emerges in different experimentally accessible geometries, while for large rotation strengths dense vortex lattices and concentric vortex ring arrangements are obtained.     
\end{abstract}

\maketitle

\section{Introduction}
Quantum and classical fluids \me{can} respond to rotation by the nucleation of vortices -- effective holes in the fluid density around which there exists a net circulation. \me{While classical fluids manifest vortices for arbitrary rotation,} vortices in quantum fluids \me{appear instead when a {\it minimum} rotation is met, and} are unique since their allowed rotational properties are restricted. Early experimental work demonstrated the feasibility of generating quantum vortices in equilibrium atomic gases \cite{matthews_1999,chevy_2000,raman_2001,abo_2001}, while ongoing work has explored non-equilibrium effects such as quantum turbulence \cite{kwon_2014}, the Berezinskii-Kosterlitz-Thouless transition \cite{hadzibabic_2006}, Kibble-Zurek dynamics \cite{sadler_2006} and the emergence of equilibrium \cite{liu_2018} and ordered states \cite{gauthier_2019}.

Degenerate quantum gases represent a powerful tool for investigating the phenomenology of analogue systems, such as the quantum simulation of effects drawn from condensed matter \cite{lewenstein_book}, quantum information and computing \cite{tomza_2019} as well as metrological applications \cite{pezze_2018}. Atomic gases benefit from being highly controllable -- here the statistics, particle interactions, dimensionality and potential landscape can be engineered with high fidelity to realize novel and exotic states of quantum matter \cite{eckardt_2017} such as the recent experimental progress with topological phenomena \cite{ozawa_2019,cooper_2019} with these systems.         
         
Quantum vortices represent the fundamental excitations of superconductors and superfluids alike, appearing in response to magnetic or rotational driving respectively. Although the atomic superfluids He II and weakly interacting Bose-Einstein condensates both nucleate vortices, the condensate fraction of He II is small (${\sim}10\%$) in comparison to that of atomic condensates which typically contain only a very small fraction of non-condensate atoms. While studying vortex physics with strongly interacting superfluids such as He II is challenging, the weakly correlated quantum fluids have shown to be strong candidates for understanding superfluid vortices and their dynamics. Experiments have generated vortex dipoles \cite{neely_2010}, observed individual vortex dynamics \cite{anderson_2000}, created vortices through interference techniques \cite{scherer_2007} and realised multiply-quantized vortices \cite{leanhardt_2002,shin_2004}. Theoretical interest in these systems has focussed on understanding the fundamental properties of the superfluid state \cite{castin_1999}, anisotropic trapping \cite{garciaripoll_2001,damski_2003,oktel_2004,watanabe_2007,mcendoo_2009,mcendoo_2010}, coherent couplings \cite{kasamatsu_2004,kasamatsu_2005,calderaro_2017,kobayashi_2019}, and the dynamics and ordering properties of few \cite{navarro_2013} and many-vortex states \cite{wilson_2015,rakonjac_2016}. 

Parallel to these developments, neutral ultracold quantum gases have become a prominent platform for the generation of synthetic forms of matter. Artificial gauge potentials represent a burgeoning sub-discipline of this field \cite{dalibard_2011,goldman_2014}, where experiments have demonstrated orbital magnetism \cite{lin_2009a,lin_2009b,lin_2011a}, spin-orbit \cite{lin_2011b} and spin-angular momentum couplings with bosons \cite{chen_2018a,chen_2018b,zhang_2019} and fermions \cite{wang_2012,cheuk_2012}. These experiments are based on versatile Raman techniques \cite{spielman_2009} that allow the internal states of atomic gases to be optically dressed to mimic the mathematical structure of a variety of gauge theories \cite{aitchison_2013}. The versatile toolbox of quantum technologies now allows for the creation of gauge potentials giving rise to spatially varying synthetic magnetic fields which  have been shown to lead to unusual manifestations of superfluidity with single component \cite{murray_2009}, spin-orbit coupled systems \cite{su_2015} and proposals for atom-surface mediated gauge theories \cite{sachdeva_2017,hejazi_2020,schloss_2020}.

All the gauge potentials realised in this way however are static -- there is no feedback between the light and the matter field. To address this, there is a growing sub-field whose aim instead is to simulate \textit{dynamical} \cite{banerjee_2012,zohar_2013,tagliacozzo_2013} gauge potentials. Then, one methodology to overcome this problem is to directly couple the gauge potential and the quantum state of the system, which naturally introduces a time-dependent feedback in the form of a \textit{density-dependent} gauge potential either in the continuum \cite{edmonds_2013a} or for lattice based theories \cite{keilmann_2011,greschner_2014,greschner_2015,jamotte_2021}. The associated phenomenology has revealed a wealth of unusual effects; in the one-dimensional context the theory violates Kohn's theorem \cite{edmonds_2014,zheng_2015} and possesses exact chiral soliton solutions \cite{aglietti_1996,dingwall_2018,dingwall_2019,ohberg_2019,bhat_2021}, as well as exhibiting unusual transport effects when confined in double well \cite{edmonds_2013b} or harmonic potentials \cite{saleh_2018}. The rotational properties of the theory present an opportunity to understand the vortex solutions and associated superfluidity in two-dimensional homogeneous \cite{butera_2016a,butera_2016b} and trapped configurations \cite{edmonds_2020}, as well as the simulation of curved spacetime for the excitations of the ground state \cite{butera_2019}. Recent work has also examined the theories mathematical structure from a hydrodynamical perspective \cite{buggy_2020a,buggy_2020b}, and proposals have now appeared that generalize the theory to support gauge theories such as those with a topological Chern-Simons structure \cite{rojas_2020}, \me{as well as density-dependent spin-orbit coupling \cite{xu_2021}.}

Complementary to its theoretical appeal, density-dependent magnetism has also been experimentally demonstrated with bosons \cite{clark_2018} and fermions \cite{gorg_2019} confined in two-dimensional optical lattices as well as for an ensemble of Rydberg atoms \cite{lienhard_2020}. \me{Very recently the first experiment in the continuum has also appeared realizing domain-walls coupled to a density-dependent gauge potential \cite{xuan_2021}.} In this work we comprehensively examine the unusual phenomenology provided by this system in a quasi-two-dimensional harmonically confined configuration, in particular focussing on the interplay of elliptical harmonic confinement and the density-dependent gauge field that manifests as a density-angular-momentum coupling, using a combination of analytical and numerical approaches to identify unconventional phenomena in experimentally accessible regimes. 

The paper is organized as follows. In Sec.~\ref{sec:model} we derive the density-dependent gauge theory and particularize the model such that the gauge potential manifests as a density-angular-momentum coupling of the atomic cloud in the quasi-two-dimensional limit. We then derive the static (Thomas-Fermi) solutions to this model, which we use to explore the stable regions of the total parameter space of the model, as well as discussing the requirements for a future experiment. Following this we present in Sec.~\ref{sec:numerics} detailed calculations of the vortical stationary states, under cylindrical and general elliptical harmonic confinement. We also explore the effect of the nonlinear rotation on the formation of vortex lattices and rings comprised of larger numbers of vortices. We summarize our finding in the Summary, Sec.~\ref{sec:summary}.                    

\section{\label{sec:model}Theoretical model}
\subsection{Density-dependent gauge theory}
\me{In what follows we demonstrate how to construct a density-dependent gauge theory using a weakly interacting two-component atomic Bose gas. This is based on the adiabatic theorem, where the gauge potentials appear as geometric vector and scaler potentials.} Our system comprises $N$ two-level atoms coupled via a coherent light-matter interaction, forming a Bose-Einstein condensate. Within the rotating-wave approximation the Hamiltonian can be written as
\begin{equation}\label{eqn:ham0}
\hat{H}=\bigg[\frac{\hat{\bf p}^2}{2m}+V({\bf r})\bigg]\otimes\mathds{1}+\hat{H}_{\rm int}({\bf r})+\hat{\mathcal{U}}_{\rm MF},
\end{equation}
where the light-matter interaction is defined as
\begin{equation}\label{eqn:lm}
\hat{\mathcal{U}}_{\rm MF}=\frac{\hbar\Omega_r}{2}\left[\begin{array}{cc}\cos\theta & e^{-i\phi}\sin\theta \\e^{i\phi}\sin\theta & -\cos\theta\end{array}\right],
\end{equation}
here $\Omega_r$ gives the strength of the light-matter coupling and $\theta$ and $\phi$ are in general spatially varying quantities. \me{While the off-diagonal components of Eq.~\eqref{eqn:lm} define the coherent coupling between the light and the matter, the diagonal terms define the detuning between the frequency of the driving laser $\omega_r$ and the atomic transition frequency $\omega_t=\omega_2-\omega_1$ between the ground and excited states, such that $\Delta=\omega_t-\omega_r$. Then one has $\Delta=\Omega_r\cos\theta$.} The other quantities that appear in Eq.~\eqref{eqn:ham0} are the two-body mean-field interactions $\hat{H}_{\rm int}=(1/2)\text{diag}[\Delta_1,\Delta_2]$, with $\Delta_j=g_{jj}n_j+g_{jk}n_k$ and $g_{jk}=4\pi\hbar^2a_{jk}/m$ defining the scattering parameter for atoms in internal states $j$ and $k$. The population density of state $j$ is $n_j=|\psi_j|^2$ and the external harmonic confinement is provided by $V({\bf r})=m(\omega_{x}^2x^2+\omega_{y}^2y^2+\omega_{z}^2z^2)/2$, and $\omega_j$ defines the trapping strength in each coordinate direction. To derive the density-dependent gauge theory, we require the eigenstates of the light-matter interaction Eq.~\eqref{eqn:lm} which for a two-level atom with internal states $\ket{1,2}$ are denoted
\begin{subequations}\label{eqn:ds}
\begin{align}
&\ket{+}=\cos\frac{\theta}{2}\ket{1}+e^{i\phi}\sin\frac{\theta}{2}\ket{2},\\
&\ket{-}=-e^{-i\phi}\sin\frac{\theta}{2}\ket{1}+\cos\frac{\theta}{2}\ket{2},
\end{align}
\end{subequations}
which obey $\hat{\mathcal{U}}_{\rm MF}\ket{\pm}=\pm\hbar\Omega_r\ket{\pm}/2$. In order to construct the density-dependent gauge theory, we use perturbation theory and Eqs.~\eqref{eqn:ds} to build \textit{perturbed dressed states}. \me{Since the motion of the atoms in the dressed states must be adiabatic, this requires that $\hbar\Omega_r\gg E_{\rm R}$, with $E_{\rm R}=p_{\rm R}^{2}/2m$ and $p_{\rm R}=\hbar k$ giving the respective atomic recoil energy and momentum \cite{cheneau_2008}. Then, the alkali-earth atoms represent a good candidate since these atoms are used for atomic metrology and in particular atomic clocks, and as such possess excited states whose lifetimes are of the order of seconds \cite{ye_2008}.} The perturbed interacting dressed basis states are defined in turn as
\begin{equation}\label{eqn:pds}
\ket{\psi_{\pm}}=\ket{\pm}\pm\frac{\Delta_d}{\hbar\Omega_r}\ket{\mp}.
\end{equation}
The mean-field dressed detuning appearing in Eq.~\eqref{eqn:pds} is defined as $\Delta_{\rm d}\me{=\langle\pm|\hat{\mathcal{U}}_{\rm MF}|\mp\rangle}=\sin\frac{\theta}{2}\cos\frac{\theta}{2}(\Delta_1-\Delta_2)/2$. In order to derive a density-dependent gauge theory, we use the perturbed dressed states, Eq.~\eqref{eqn:pds} to define a state $\ket{\chi}=\sum_{j=+,-}\Psi_j({\bf r},t)\ket{\psi_j}$ along with Eq.~\eqref{eqn:ham0}, and since the qualitative details \me{of the resulting physics} do not depend on which of the two dressed states we choose \me{to project into}, the atomic motion will be projected into the $\ket{\psi_+}$ state. Hence the effective Hamiltonian is written
\begin{equation}\label{eqn:hamp}
\hat{H}_+=\frac{({\bf p}-{\bf A}_+)^2}{2m}+W_{+}+\frac{\hbar\Omega_r}{2}+\Delta_{+}+V({\bf r}),
\end{equation}
here the dressed mean-field atomic interactions appearing in Eq.~\eqref{eqn:hamp} are written as $\Delta_{+}\me{=\langle+|\hat{\mathcal{U}}_{\rm MF}|+\rangle}=(\Delta_1\cos^2\frac{\theta}{2}+\Delta_2\sin^2\frac{\theta}{2})/2$, while the two \me{arising} geometric potentials are consequences of the adiabatic atomic motion. The vector potential is defined by ${\bf A}_{+}=i\hbar\langle\psi_{+}|\nabla\psi_{+}\rangle$, while the scalar potential is $W_{+}=\hbar^2|\langle\psi_+|\nabla\psi_-\rangle|^2/2m$. For atomic motion in the $\ket{\psi_{+}}$ state these two geometric potentials are
\begin{subequations}
\begin{align}\label{eqn:gvec}
{\bf A}_{+}=&-\frac{\hbar}{2}(1-\cos\theta)\nabla\phi+\frac{\Delta_{\rm d}}{\Omega_r}\nabla\phi\sin\theta,\\\nonumber
W_{+}=&\frac{\hbar^2}{8m}(\nabla\theta)^2+\frac{\hbar^2}{8m}\sin^2\theta(\nabla\phi)^2\\&+\frac{\hbar}{2m}\frac{\Delta_{\rm d}}{\Omega_r}\sin\theta\cos\theta(\nabla\phi)^2-\hbar\nabla\theta\cdot\nabla\frac{\Delta_{\rm d}}{\Omega_r}.\label{eqn:gpot}
\end{align}
\end{subequations}
Then in order to derive a mean-field equation of motion for the atoms, we write the Dirac-Frenkel action $\mathcal{S}$ as
\begin{equation}
\mathcal{S}[\Psi_{+}^{*},\Psi_{+}]{=}\int dt\int d{\bf r}\Psi_{+}^{*}({\bf r},t)\bigg[i\hbar\frac{\partial}{\partial t}{-}\hat{H}_{+}\bigg]\Psi_{+}({\bf r},t)
\end{equation}
which can in turn be extremized by computing $\delta\mathcal{S}/\delta\Psi_{+}^{*}=0$, from which one obtains the generalized Gross-Pitaevskii equation \cite{edmonds_2013a,butera_2017}
\begin{align}\nonumber
&i\hbar\frac{\partial}{\partial t}\Psi_{+}{=}\bigg[\frac{({\bf p}{-}{\bf A}_{+})^2}{2m}{+}W_{+}{+}{\bf a}_1\cdot{\bf j}{+}\frac{\hbar\Omega_r}{2}{+}2\Delta_{+}{+}V({\bf r})\bigg]\Psi_{+}\\&+\bigg[n_+\bigg(\frac{\partial W_+}{\partial\Psi_{+}^{*}}{-}\nabla\cdot\frac{\partial W_{+}}{\partial\nabla\Psi_{+}^{*}}\bigg){-}\frac{\partial W_+}{\partial\nabla\Psi_{+}^{*}}\cdot\nabla n_{+}\bigg]\label{eqn:ggpe}
\end{align} 
for atomic motion in the $\ket{\psi_+}$ dressed state. Here the strength of coupling to the gauge potential is ${\bf a}_1=\nabla\phi\Delta_{\rm d}\sin\theta/n_{+}\Omega_r$. The generalised Gross-Pitaevskii equation \eqref{eqn:ggpe} includes a number of additional terms arising from the density-dependence of the geometric potentials, \eqref{eqn:gvec} and \eqref{eqn:gpot}, in-particular the current nonlinearity
\begin{equation}\label{eqn:cur}
{\bf j}=\frac{\hbar}{2mi}\bigg[\Psi_{+}\bigg(\nabla+\frac{i}{\hbar}{\bf A}_{+}\bigg)\Psi_{+}^{*}-\Psi_{+}^{*}\bigg(\nabla-\frac{i}{\hbar}{\bf A}_{+}\bigg)\Psi_{+}\bigg].
\end{equation}
The theory supports two small parameters -- $\theta=\Omega_r/\Delta$, which gives the ratio of the Rabi frequency to the detuning, and secondly $\varepsilon=n(g_{11}-g_{22})/4\hbar\Delta$ that underpins the collisional and coherent interactions. Since both of these parameters are assumed to be small, we can expand Eqs.~\eqref{eqn:gvec}-\eqref{eqn:gpot} to first order in $\varepsilon$ and $\theta$, which gives the following simplified relations for the vector potential
\begin{equation}\label{eqn:veca}
{\bf A}_+=-\frac{\hbar\theta^2}{4}[1-4\varepsilon]
\end{equation}  
and the scalar potential
\begin{figure}[t]
\includegraphics[width=\columnwidth]{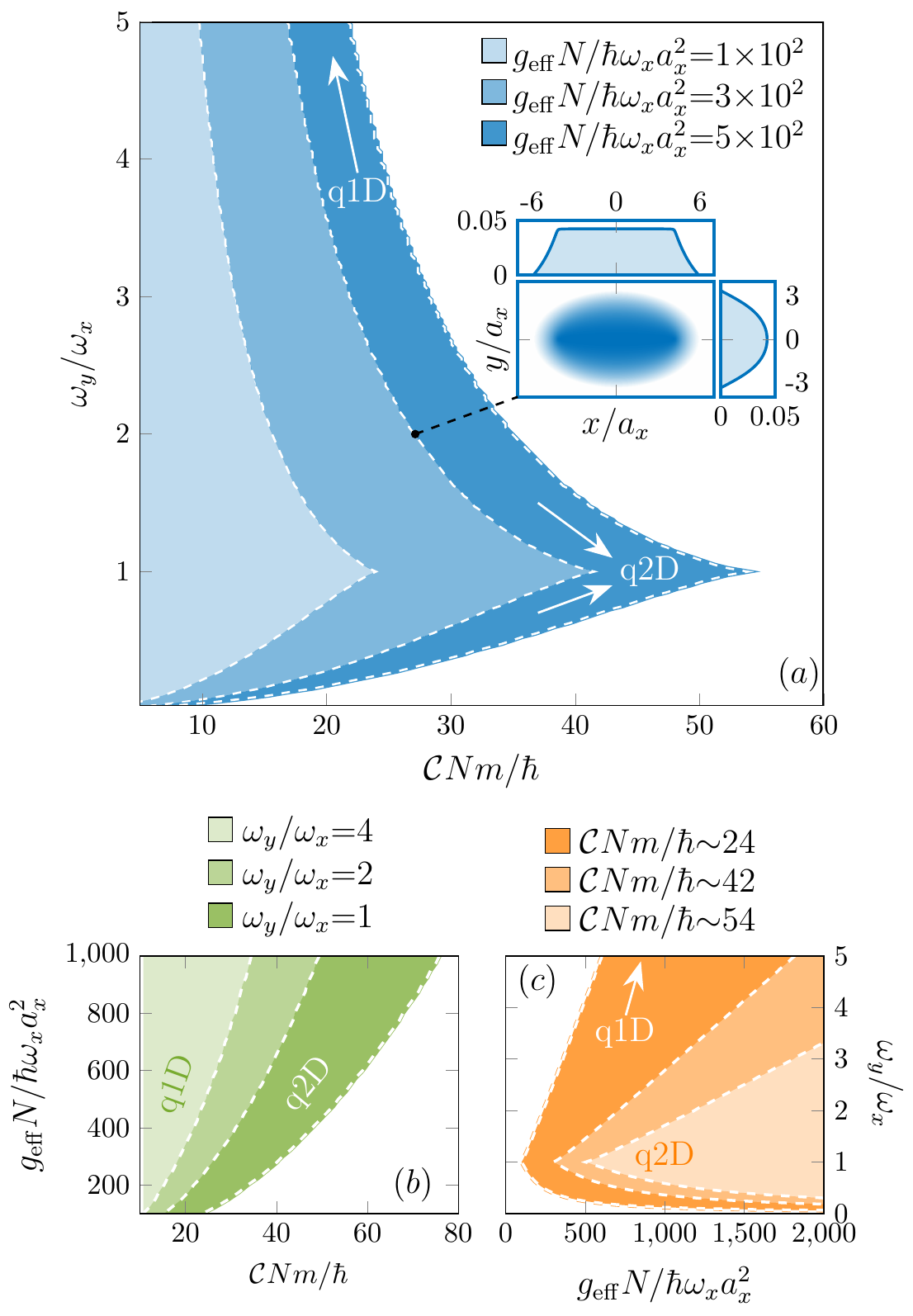}
\caption{\label{fig:stab}(color online) Nonlinear rotation stability. The boundaries (maximum nonlinear rotation strength $\mathcal{C}$) are shown for various fixed interaction strengths $g_{\rm eff}$ in the $\omega_y/\omega_x$ and $\mathcal{C}$ parameter space in panel (a), along with an example inset density profile $|\psi|^2$ for $(\mathcal{C}Nm/\hbar{\sim}27,\omega_y/\omega_x{=}2)$. Panels (b) and (c) show the counterparts of panel (a) except with fixed trap ellipticity and nonlinear rotation strength respectively.}
\end{figure}
\begin{equation}
W_+{=}\frac{\hbar^2}{2}\bigg[\frac{(\nabla\theta)^2[1{-}4\varepsilon]{+}\theta^2(\nabla\phi)^2[1{+}4\varepsilon]}{4m}{-}\nabla\theta^2\cdot\nabla\varepsilon\bigg]
\end{equation}  
which together with Eqs.~\eqref{eqn:ggpe}-\eqref{eqn:cur} can then be used to construct a simplified density-dependent gauge theory, where the equation of motion for the condensate becomes
\begin{equation}\label{eqn:nlr}
i\hbar\frac{\partial\psi}{\partial t}=\bigg[-\frac{\hbar^2}{2m}\nabla^2+V({\bf r})-\Omega({\bf r},t)\hat{L}_z+g_{\rm eff}n\bigg]\psi.
\end{equation}
Equation \eqref{eqn:nlr} defines a type of generalized nonlinear Schr\"odinger equation where the condensate experiences a {\it nonlinear} as well as rigid body rotation through the density-dependent rotation frequency $\Omega({\bf r},t)=\Omega+\mathcal{C}n({\bf r},t)$. \me{The two geometric potentials $\theta$ and $\phi$ are given by $\theta=\theta_0\rho$ and $\phi=+\varphi$ respectively, where $\theta_0$ is a constant of proportionality arising from assuming the angle $\theta$ is small. Then, the} effective strength of the nonlinear rotation is $\mathcal{C}=\theta_{0}^2(g_{11}-g_{22})/2m\Omega_r$, while $g_{\rm eff}=g_{\rm 11}+\hbar\theta_{0}^{2}(g_{11}-g_{22})/m\Omega_r$ defines the effective two-body scattering parameter. Finally, the angular momentum operator is given by $\hat{L}_z=\boldsymbol{\rho}\times{\bf p}$. In what follows we specialize to quasi-two-dimensional harmonic confinement so that the atomic cloud adopts a pancake geometry with $\omega_z\gg\omega_{x,y}$. This in turn permits us to write the full three-dimensional mean-field wave function in the factorized form $\psi({\bf r},t)=\psi(x,y,t)\exp(-z^2/2a_{z}^2)/\sqrt[4]{\pi a_{z}^2}$ which can be combined with Eq.~\eqref{eqn:nlr} to project out the dynamics of the $z$-coordinate. This causes the effective strengths of the two nonlinear terms $g_{\rm eff}$ and $\mathcal{C}$ to be scaled by the factor $1/\sqrt{2\pi}a_z$, which in what follows will be absorbed into the definitions of these two parameters for notational convenience. 

\subsection{\label{sec:exp}Experimental considerations}

Given the active experimental interest in these types of synthetic gauge theories \me{\cite{clark_2018,gorg_2019,lienhard_2020,xuan_2021}}, let us outline the key ingredients required to realize the density-angular-momentum coupled gauge theory, Eq.~\eqref{eqn:nlr}. The geometric phase methodology on which the model is based is generated by the light-matter interaction, Eq.~\eqref{eqn:lm}. Laser light possessing fixed angular momentum can be exploited in order to generate the required mathematical structure of the gauge theory \cite{juzeliunas_2005}. The recent experiments demonstrating spin-angular-momentum-coupled Bose-Einstein condensates used Laguerre-Gaussian laser light with a radially varying electric field profile \cite{chen_2018a,chen_2018b,zhang_2019}
\begin{equation}\label{eqn:ef}
\mathcal{E}({\bf r})=\mathcal{E}_0e^{-i\ell\varphi}\bigg(\frac{\rho}{w}\bigg)^{|\ell|}e^{-\rho^2/w^2}e^{ikz},
%I(\rho)=I_{0}\bigg(\frac{\rho}{w}\bigg)^{2|\ell|}e^{-2\rho^2/w^2},
\end{equation}
where $\mathcal{E}_0$ is the amplitude, $w$ is waist of the beam and $k$ is the wavenumber. For $\ell=+1$, Eq.~\eqref{eqn:ef} has precisely the spatial structure that is required since $\phi=+\varphi$ and \me{$\theta\propto\rho$}. %As well as needing the correct spatially structured light, 
Accompanying these requirements the perturbative requirement must also be satisfied such that $\varepsilon\ll1$, and since $\varepsilon$ directly depends on the difference of scattering lengths $g_{11}-g_{22}$ (see Eq.~\eqref{eqn:veca}) optical Feshbach resonances for alkali-earth atoms represent an important resource for tuning this parameter \cite{enomoto_2008}.

\subsection{Static solutions and stability analysis}

The energy functional associated with eq.~\eqref{eqn:nlr} can be written in a hydrodynamic picture described by the density $n(\rho,\varphi)$ and phase $\vartheta(\rho,\varphi)$ degrees of freedom where $\rho,\varphi$ define the polar coordinates using the Madelung decomposition $\psi(\rho,\varphi)=\sqrt{n(\rho,\varphi)}\exp(i\vartheta(\rho,\varphi))$ as 
\begin{align}\nonumber
E[n,\vartheta]=&\int d{\boldsymbol{\rho}} n\bigg[\frac{\hbar^2}{2m}\frac{|\nabla n|^2}{4n^2}+\frac{m}{2}{\bf v}^2+\frac{i\hbar}{2mn}\nabla n\cdot\mathcal{A}\\&-\frac{\mathcal{A}^2}{2m}+V(\rho)+\frac{g_{\rm eff}n}{2}-\mu'\bigg],\label{eqn:hen}
\end{align}
where ${\bf v}=(\hbar\nabla\vartheta-\mathcal{A})/m$ defines the hydrodynamic kinetic velocity, while $\mathcal{A}=m\boldsymbol{\Omega}\times\boldsymbol{\rho}$ and $\boldsymbol{\Omega}=\hat{e}_z(\Omega+\mathcal{C}n/2)$. Working in the limit \me{$g_{\rm eff}N/\hbar\omega_x a_{x}^2\gg1$} allows us to drop the quantum pressure term appearing in Eq.~\eqref{eqn:hen}. Then by first minimizing the energy functional Eq.~\eqref{eqn:hen} with respect to the hydrodynamic phase $\vartheta(\rho,t)$ the superfluid velocity ${\bf v}_{\rm sf}=\boldsymbol{\Omega}\times\boldsymbol{\rho}$ is obtained. Subsequently inserting ${\bf v}_{\rm sf}$ into  Eq.~\eqref{eqn:hen} and then minimizing with respect to the hydrodynamic density $n(\rho,\varphi)$ we obtain the rotating frame Thomas-Fermi distribution
\begin{equation}\label{eqn:rtf}
V(\rho)-\frac{m}{2}\rho^2\bigg(\Omega^2+2\Omega\mathcal{C}n+\frac{3}{4}\mathcal{C}^2n^2\bigg)+g_{\rm eff}n=\mu',
\end{equation}
here $\rho^2=x^2+y^2$ and the elliptical harmonic trap is $V(\rho)=\frac{1}{2}m(\omega_{x}^2x^2+\omega_{y}^2y^2)$, while $\mu'$ is the chemical potential in the rotating frame and the cartesian boundaries of the cloud are found from $R_{x,y}^2=2\mu'/m(\omega_{x,y}^2-\Omega^2)$. In the limit $\mathcal{C}\rightarrow0$ one recovers the standard rotating frame Thomas-Fermi distribution with $n(\rho)=(\mu'-[V(\rho)-\tfrac{1}{2}m\Omega^2\rho^2])/g_{\rm eff}$. By imposing the normalization condition
\begin{equation}\label{eqn:norm}
\int d\boldsymbol{\rho} \theta(\mu'-V(\rho)) n(\rho)=N,
\end{equation}
in this limit we obtain 
\begin{figure*}[t]
\includegraphics[width=0.95\textwidth]{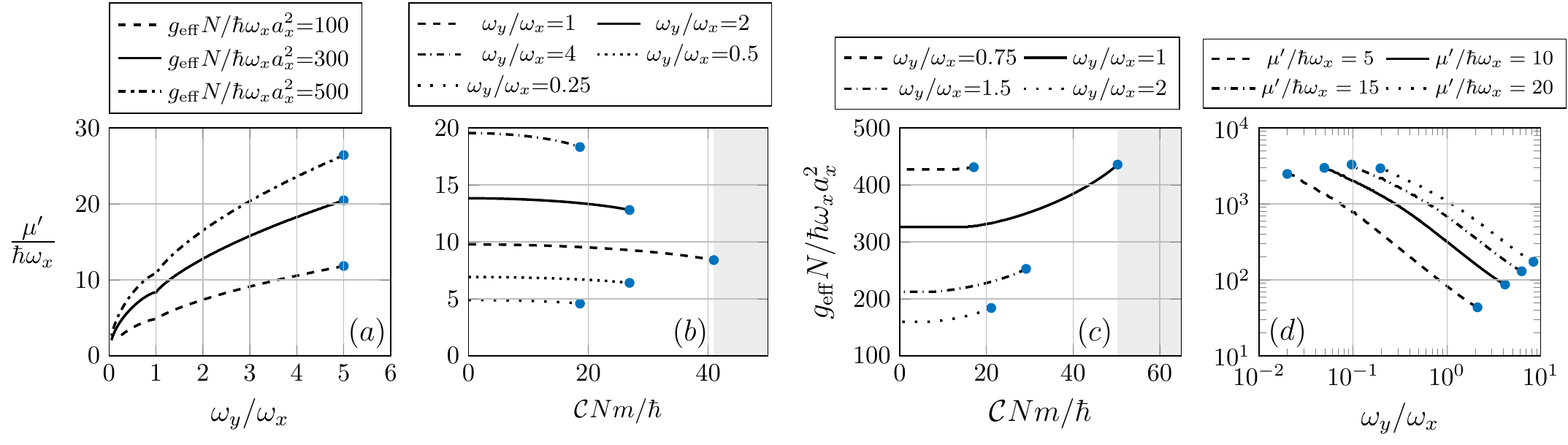}
\caption{\label{fig:mu}(color online) Fixed parameter solutions to Eq.~\eqref{eqn:mun}. Panels (a) and (b) show the chemical potential $\mu'$ for fixed interaction strength $g_{\rm eff}$ and trap anisotropy $\omega_y/\omega_x$ respectively. Panel (c) depicts the allowed solutions for fixed $\mu'/\hbar\omega_x=10$ and $\omega_y/\omega_x$ obtained from Eq.~\eqref{eqn:mun} in the $(g_{\rm eff},\mathcal{C})$ parameter space, while (d) shows allowed solutions for fixed $\mu'$ in the $(g_{\rm eff},\omega_y)$ parameter space with $\mathcal{C}Nm/\hbar=10$. The grey shaded regions in (b) and (c) indicate the boundary to maximum nonlinear rotation for $\omega_y=\omega_x$.}
\end{figure*}
\begin{equation}\label{eqn:muz}
\mu'=\sqrt{\frac{mg_{\rm eff}N}{\pi}}\sqrt[4]{(\omega_{x}^2-\Omega^2)(\omega_{y}^2-\Omega^2)} 
\end{equation}
which defines the corresponding chemical potential, which can be shown to possesses real ($\mu'>0$) solutions when $\Omega<\omega_y$ and $\Omega<\omega_x$. \me{The two trapping frequencies $\omega_{x,y}$ define the maximum rigid body rotation strength.} Then, Eq.~\eqref{eqn:rtf} is quadratic in the density distribution $n(\rho,\varphi)$ and can be solved straight-forwardly to give
\begin{widetext}
\begin{equation}\label{eqn:tfexp}
n(x,y)=-\frac{4}{3m\mathcal{C}^2\rho^2}\bigg\{-\big[g_{\rm eff}-m\Omega\mathcal{C}\rho^2\big]+\sqrt{\big[g_{\rm eff}-m\Omega\mathcal{C}\rho^2\big]^2+\frac{3m\mathcal{C}^2\rho^2}{2}\big[\tfrac{1}{2}m([\omega_{x}^2-\Omega^2]x^2+[\omega_{y}^2-\Omega^2]y^2)-\mu'\big]}\bigg\},
\end{equation}
\end{widetext}
Equation \eqref{eqn:tfexp} gives in general the shape of the background of the trapped condensate under both nonlinear and rigid body rotation. Here we will consider the limit $\Omega=0$ which corresponds to a vortex-free state, and allows us to construct semi-analytic and analytical solutions. The solutions in general are bounded by the ellipse $R_{x}^{2}(\omega_{x}^2-\Omega^2)+R_{y}^{2}(\omega_{y}^2-\Omega^2)=2\mu'/m$. The normalization integral Eq.\eqref{eqn:norm} can then be partially evaluated by switching to polar coordinates, and leads to the expression
\begin{widetext}
\begin{equation}\label{eqn:mun}
\text{sinh}^{-1}\sqrt{\frac{\omega_{x}^2(1{-}\alpha^2)}{\omega_{y}^2{-}\omega_{x}^2}}{-}\text{sinh}^{-1}\sqrt{\frac{\omega_{x}^{2}}{\omega_{y}^2{-}\omega_{x}^2}}{+}\int\limits_{0}^{2\pi} \frac{d\varphi}{2\pi}\frac{\omega_{x}\alpha}{\sqrt{\omega_{x}^{2}{+}[\omega_{y}^2{-}\omega_{x}^2]\sin^2\varphi}}\text{tanh}^{-1}\bigg[\frac{\omega_{x}\alpha}{\sqrt{\omega_{x}^{2}{+}[\omega_{y}^2{-}\omega_{x}^2]\sin^2\varphi}}\bigg]{=}\frac{3m\mathcal{C}^2N}{8\pi g_{\rm eff}},
\end{equation}
\end{widetext}
where for brevity we have defined the dimensionless ratio $\alpha=\sqrt{3}\mathcal{C}\mu'/2\omega_{x}g_{\rm eff}$ which connects the chemical potential $\mu'$, nonlinear rotation strength $\mathcal{C}$ and quasi-two-dimensional scattering parameter $g_{\rm eff}$. The final angular integration appearing in Eq.\eqref{eqn:mun} cannot in general be reduced to an elementary function, except for the special case of when the trap has cylindrical symmetry so $\omega_y=\omega_x$. This leads to the expression
\begin{equation}\label{eqn:mus}
\ln\bigg[1{-}\frac{3\mathcal{C}^2{\mu'}^2}{4\omega_{x}^{2}g_{\rm eff}^{2}}\bigg]{+}\frac{\sqrt{3}\mathcal{C}\mu'}{\omega_xg_{\rm eff}}\tanh^{-1}\bigg[\frac{\sqrt{3}\mathcal{C}\mu'}{2\omega_xg_{\rm eff}}\bigg]{=}\frac{3m\mathcal{C}^2N}{4\pi g_{\rm eff}}.
\end{equation}
Equation \eqref{eqn:mus} defines an implicit relationship between the chemical potential and the various interaction parameters, whose logarithmic nature means that the allowed solutions exist on the finite domain $-2\omega_x g_{\rm eff}/\sqrt{3}\mu'\leq\mathcal{C} \leq2\omega_x g_{\rm eff}/\sqrt{3}\mu'$. From this one can identify a minimum value of the chemical potential $\mu_{\rm min}'=2\omega_xg_{\rm eff}/\sqrt{3}\mathcal{C}_{\rm max}$, which using Eq.~\eqref{eqn:mus} in turn allows us to obtain the maximum nonlinear rotation strength $\mathcal{C}_{\rm max}$ as
\begin{figure*}[t]
\includegraphics[width=0.9\textwidth]{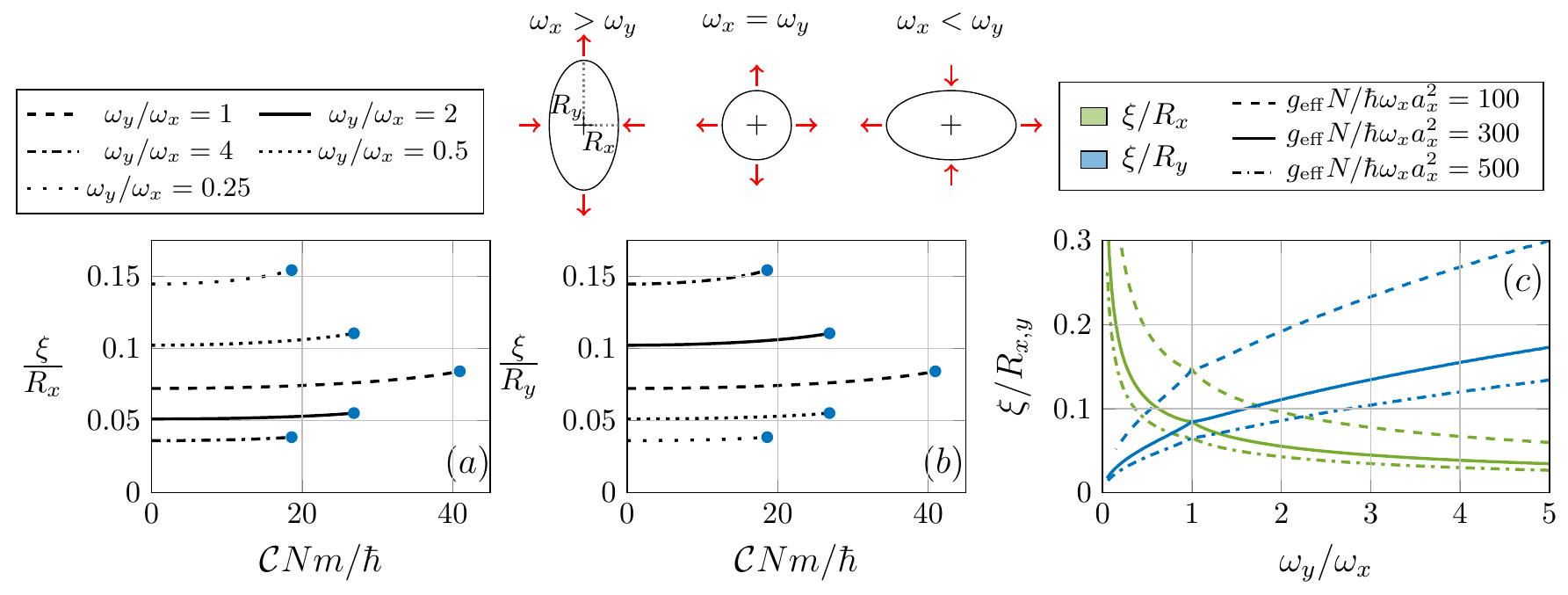}
\me{\caption{\label{fig:lda}(color online) Local density approximation. The ratio $\xi/R_{x,y}$ is shown in panels (a) and (b) with constant $g_{\rm eff}N/\hbar\omega_x a_{x}^2=300$ throughout for various fixed $\omega_y/\omega_x$ as a function of $\mathcal{C}$, while (c) shows $\xi/R_{x,y}$ as a function of $\omega_y/\omega_x$. The three schematic diagrams above (c) show the geometry of the cloud for $\omega_y/\omega_x<1,1>1$.}}
\end{figure*}
\begin{equation}\label{eqn:cm}
\mathcal{C}_{\rm max}=\sqrt{\frac{8\pi\ln2}{3}\frac{g_{\rm eff}}{Nm}},
\end{equation}
which for $\mathcal{C}>\mathcal{C}_{\rm max}$ (or $\mathcal{C}<-\mathcal{C}_{\rm max}$) means the nonlinear rotation overcomes the harmonic confinement, and the condensate no longer exists, a situation analogous to the case of pure ($\mathcal{C}=0$) rigid body rotation when the trapped gas is rotated faster than harmonic trapping strength such that $\Omega>\omega_{x,y}$ (viz. Eq.~\eqref{eqn:muz}). Using Eq.~\eqref{eqn:cm} we obtain the corresponding minimum chemical potential as 
\begin{equation}\label{eqn:mutf}
\mu_{\rm min}'=\sqrt{\frac{N}{\pi\ln4}m\omega_{x}^2g_{\rm eff}}.
\end{equation}
Since the chemical potential can be related to the total energy via the standrd thermodynamic relation $\mu_{\rm min}'=\partial E_{\rm min}/\partial N$, Eq.~\eqref{eqn:mutf} yields the relation $E_{\rm min}=(2/3)N\mu_{\rm min}'$. Then, using the definition $R_{\rm TF}^2=2\mu_{\rm min}'/m\omega_{x}^2$ we obtain the Thomas-Fermi length scale associated with Eqs.~\eqref{eqn:cm}-\eqref{eqn:mutf}
\begin{equation}\label{eqn:tfr}
R_{\rm TF}=\bigg(\frac{2}{\pi\ln2}\frac{g_{\rm eff}N}{m\omega_{x}^2}\bigg)^{1/4}.
\end{equation}
Knowledge of the chemical potential given by Eq.~\eqref{eqn:mutf} and intrinsic length scale Eq.~\eqref{eqn:tfr} provide useful insight for potential future experimental studies, since these points in the parameter space show predominantly the effect of the nonlinear rotation as well as providing key information about the stable regions of the total parameter space of the model. In general for elliptical harmonic confinement ($\omega_y\neq\omega_x$) Eq.~\eqref{eqn:mun} can be solved numerically to understand the regions where the condensate exists, and as such the maximum nonlinear rotation strength for the cylindrically symmetry, Eq.~\eqref{eqn:cm} provides a useful reference point to understand the more general elliptical case ($\omega_y\neq\omega_x$).

Figure \ref{fig:stab} shows the numerical solutions obtained from Eq.~\eqref{eqn:mun} in the parameter space of the trap anisotropy $\omega_y/\omega_x$, nonlinear rotation strength $\mathcal{C}Nm/\hbar$ and two-body mean-field interaction strength $g_{\rm eff}N/\hbar\omega_xa_{x}^2$. The allowed solutions are presented in the $(\omega_y/\omega_x,\mathcal{C}Nm/\hbar)$ parameter space in panel (a). Here, each of the three shaded regions corresponds to the allowed solutions for a particular fixed value of the two-body mean-field strength $g_{\rm eff}N/\hbar\omega_xa_{x}^2=1\times10^2,3\times10^2,5\times10^2$. The dotted white boundaries in each case represent the maximum value of the nonlinear rotation strength $\mathcal{C}Nm/\hbar$ at which a solution can be obtained for the Thomas-Fermi density Eq.~\eqref{eqn:tfexp}. The accompanying inset shows an example solution for $|\psi|^2$, for $g_{\rm eff}N/\hbar\omega_x a_{x}^2=3\times10^2$ with trap anisotropy $\omega_y/\omega_x=2$. The two panels above and below this show cross-sections of $|\psi(x,y=0)|^2$ and $|\psi(x=0,y)|^2$ respectively. The density cut along $y=0$ reveals the effect of the density-angular-momentum coupling, giving a flattened top to density profile, similar to a quantum droplet \cite{tengstrand_2019}. Then, the lower row of Fig.~\ref{fig:stab} \me{shows} a pair of panels depicting the allowed solutions for fixed trap anisotropy in (b) and fixed nonlinear rotation strength in (c). Note that the white-dashed boundaries shown in panel (c) depict the {\it minimum} value of the two-body mean-field interaction $g_{\rm eff}$ at which a solution to Eqs.~\eqref{eqn:tfexp} and \eqref{eqn:mun} can be obtained. For the case of cylindrical symmetry, this minimum can be shown to be 
\begin{figure*}[t]
\includegraphics[scale=0.88]{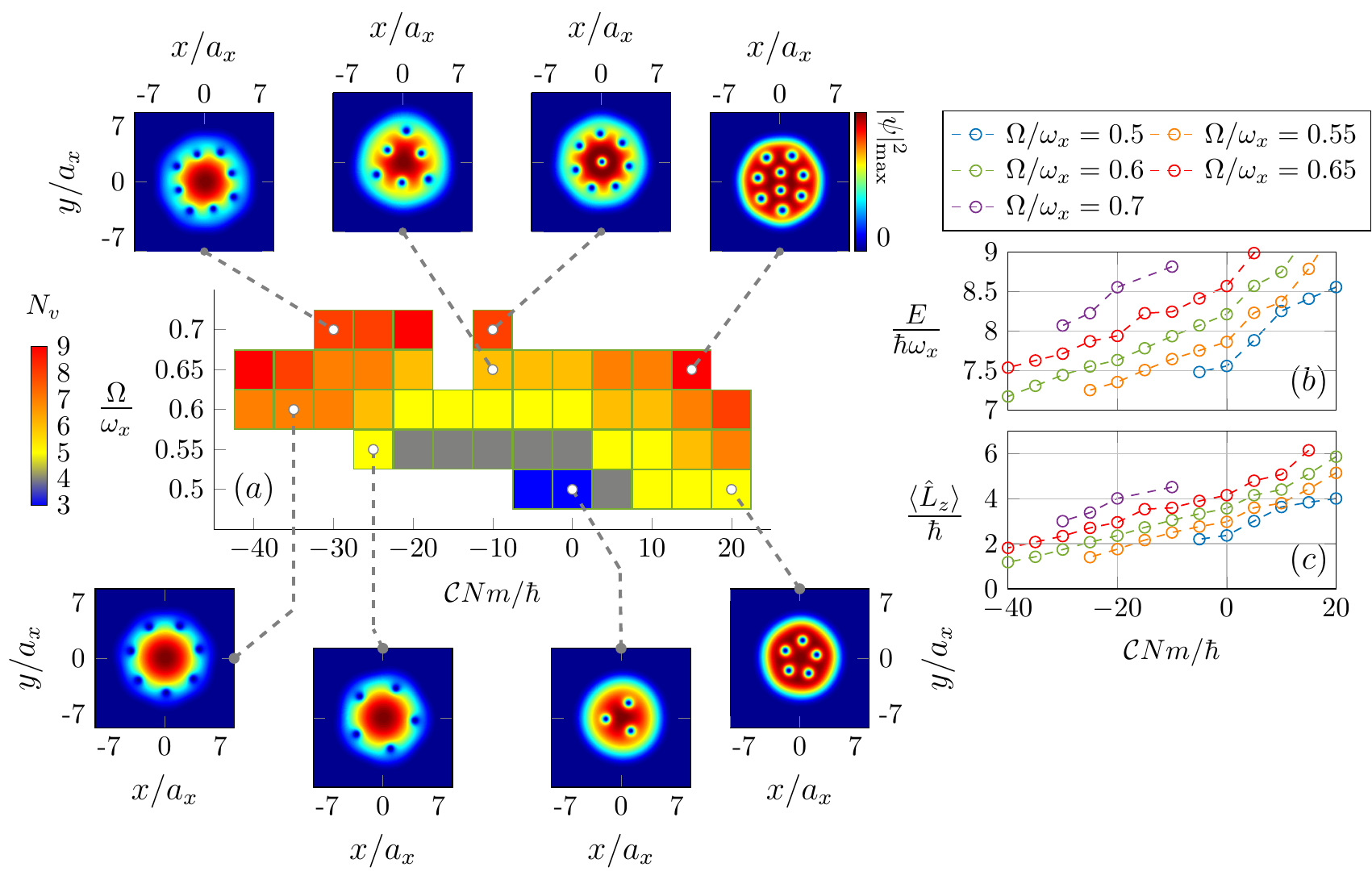}
\caption{\label{fig:vortnum}(color online) Vortex ground state diagram. Panel (a) shows the number of vortices $N_v$ as a function of the rigid body $\Omega$ and nonlinear rotation strengths $\mathcal{C}$ for forty-four individual ground states. Then example ground states $|\psi(x,y)|^2$ are shown for different choices of $(\Omega,\mathcal{C})$, indicated by the dashed grey lines. The energy $E$ computed using Eq.~\eqref{eqn:energy} and angular momentum $\langle\hat{L}_z\rangle$ corresponding to the data presented in (a) are shown in panels (b) and (c) respectively.}
\end{figure*}
\begin{equation}
g_{\rm eff}^{\rm min}=\frac{3}{8\pi\ln2}\mathcal{C}^2Nm. 
\end{equation}
Then in (a)-(c) the quasi-two and quasi-one-dimensional regions are annotated with accompanying arrows. \me{The addition of the nonlinear rotation means we must also consider the effect of the parameter $\mathcal{C}$ on the validity of the Thomas-Fermi approximation. Since we work in a perturbative regime $\Delta_{\rm d}\ll\hbar\Omega_r$, the effect of the gauge field might either be negligible, or require very large coupling strengths to have any effect that would violate perturbation theory. From the analysis presented in Fig.~\ref{fig:stab}-\ref{fig:lda} for typical two-body interaction strengths of order $g_{\rm eff}N/\hbar\omega_x a_{x}^2\sim 10^2$ it is already seen that nonlinear rotation strengths of the order $\mathcal{C}Nm/\hbar\sim10$ are sufficient to observe effects stemming from the nonlinear rotation, which does not violate the assumption $\Delta_{\rm d}\ll\hbar\Omega_r$.}

Figure \ref{fig:mu} shows example solutions for different fixed parameters. Panel (a) shows example chemical potentials $\mu'/\hbar\omega_x$ calculated as a function of the trap anisotropy for three fixed individual values of the two-body mean-field interaction strength $g_{\rm eff}N/\hbar\omega_xa_{x}^2=1\times10^2,3\times10^2,5\times10^5$. We observe that the point at which the solutions terminate shifts to larger values of $\mu'/\hbar\omega_x$ as the two-body mean-field strength is increased. Then, panel (b) shows the chemical potential as a function of the nonlinear \me{rotation} strength, for different fixed values of the trap anisotropy with $g_{\rm eff}N/\hbar\omega_xa_{x}^2=3\times10^2$. Here the \me{nonlinear rotation strength $\mathcal{C}$} attains a maximum value for the case of cylindrical symmetry ($\omega_y=\omega_x$) and is found to \me{possess a maximum value that is always smaller than this for} $\omega_y>\omega_x$ and $\omega_y<\omega_x$, i.e. as the cloud goes from a quasi-two to quasi-one-dimensional geometry. Then, panel (c) depicts four example solutions with $\omega_y/\omega_x=0.75,1,1.5,2$ in the $(g_{\rm eff},\mathcal{C})$ parameter space with constant $\mu'/\hbar\omega_x=10$. Again the point of cylindrical symmetry ($\omega_y=\omega_x$) represents the point with the largest range of solutions for $\mathcal{C}$, while either reducing or increasing the anisotropy of the harmonic trap always reduces the range of available solutions, similar to (b). Meanwhile (d) shows example solutions for fixed $\mu'/\hbar\omega_x=10$ and $\mathcal{C}Nm/\hbar=10$ in the $(g_{\rm eff},\omega_y/\omega_x)$ parameter space. Plotted on a log-log scale these solutions are revealed to be quasi-linear, with a maximum and minimum allowed solution for a given fixed pair of parameters. Increasing the chemical potential $\mu'$ has the effect of shifting the solutions to higher $g_{\rm eff}$, as well as simultaneously increasing the overall observed range of $\omega_y/\omega_x$ for a given pair of parameters.

\me{It is also also important to consider the validity of the theory in terms of the local density approximation. On physical grounds this amounts to the size of an individual vortex core being much smaller than the size of the condensate, consequentially the healing length $\xi=\hbar/\sqrt{m\mu'}$ of the vortex must be much smaller than the Thomas-Fermi length scales $R_{x,y}$ or
\begin{equation}\label{eqn:lda}
\frac{\xi}{R_{x,y}}=
\begin{cases}
\dfrac{\hbar\omega_x}{\sqrt{2}\mu'}\ll1\ (x),\\
\dfrac{\hbar\omega_x}{\sqrt{2}\mu'}\dfrac{\omega_y}{\omega_x}\ll1\ (y),
\end{cases}
\end{equation}
Figure \ref{fig:lda} presents numerical calculations of the ratio $\xi/R_{x,y}$. Panels (a) and (b) show $\xi/R_x$ and $\xi/R_y$ respectively for fixed two-body interaction strength $g_{\rm eff}N/\hbar\omega_x a_{x}^2=300$, for various fixed trap anisotropies $\omega_y/\omega_x=\tfrac{1}{4},\tfrac{1}{2},1,2,4$. In all cases Eq.~\eqref{eqn:lda} is satisfied. Panel (c) shows instead $\xi/R_{x,y}$ computed as a function of $\omega_y/\omega_x$, for various fixed two-body interaction strengths. The data for $\xi/R_x$ (green) is monotonically decreasing on the interval $0<\omega_y/\omega_x\leq 5$, whereas the data for $\xi/R_y$ (blue) monotonically increases on this interval. This can be interpreted in a straightforward way -- for $\omega_y/\omega_x<1$ the data for $\xi/R_x$ shows the local density approximation is not as good as for $\omega_y/\omega_x>1$. This situation is reversed for the equivalent $\xi/R_y$ data, where the local density approximation is instead improving for $\omega_y/\omega_x<1$. Finally the three schematic diagrams above panel (b) illustrate the three situations $\omega_y/\omega_x<1,1,>1$. For the case of cylindrical symmetry we can write down an exact form of Eq.~\eqref{eqn:lda} using Eqs.~\eqref{eqn:mutf} and \eqref{eqn:tfr}, giving  
\begin{equation}
\sqrt{\frac{\pi\ln2}{N}}\frac{\hbar}{\sqrt{mg_{\rm eff}}}\ll1.
\end{equation} 
Finally, let us comment on the presence of the cusps that appear in Figs.~\ref{fig:stab}(a) and (c), Fig.~\ref{fig:mu}(a) and also Fig.~\ref{fig:lda}(c) at the symmetry point $\omega_y=\omega_x$. Examining the mathematical structure of Eq.~\eqref{eqn:mun} we note that the first two terms on the left hand side are a difference of inverse hyperbolic sines whose arguments involve square roots of the function $(\omega_{y}^2/\omega_{x}^2-1)^{-1}$ which is sharply peaked around the symmetry point $\omega_y=\omega_x$. The difference of these two functions (the argument of the second is shifted with respect to first by a factor of $1-\alpha^2$, which effectively removes the singularity at $\omega_y=\omega_x$) can be understood as the origin of these cusps. %We can interpret the appearance of these cusps as a mathematical consequence of the effective projection of the total parameter space ($\mu',\mathcal{C},g_{\rm eff},\omega_{y,x}$) into a particular subset of these parameters, for example in Fig.~\ref{fig:stab}(a) the chemical potential $\mu'$ is projected out for constant $g_{\rm eff}$ into the space ($\mathcal{C},\omega_y/\omega_x$). These cusps arise from the projection of some non-trivial function, in this case Eq.~\eqref{eqn:mun} containing at least a single pair of maxima and minima (a fold) onto a plane, i.e. a subset of the total parameter space \cite{acheson_book}. 
As the solutions are continuous through the symmetry point in these examples, these cusps can be understood as an interesting mathematical artifact of Eq.~\eqref{eqn:mun}, rather than having any measurable physical consequence in a potential experimental.
\subsection{Existence of stationary solutions}
Since this model (Eq.~\eqref{eqn:nlr}) possesses an unusual nonlinear structure, it is important to consider the nature of the nonlinear solutions, and in-particular their dynamical behaviour. Since the model is Hermitian, the dynamical evolution of solutions obtained from the regions of the parameter space where they exist are stationary with a well defined energy. An explicit way to see this is to switch to the hydrodynamic picture using the Madelung transformation $\psi({\bf r},t)=\sqrt{n({\bf r},t)}\exp(i\vartheta({\bf r},t))$, giving 
\begin{equation}\label{eqn:cl}
\frac{\partial n({\bf r})}{\partial t}+\underline{\nabla}\cdot\bigg(n({\bf r})\bigg\{{\bf v}_{\vartheta}-\frac{1}{2}\big[\boldsymbol{\Omega}({\bf r},t)+\boldsymbol{\Omega}\big]\times{\bf r}\bigg\}\bigg)=0,
\end{equation}
where ${\bf v}_{\varphi}=\hbar\underline{\nabla}\vartheta/m$, $\boldsymbol{\Omega}({\bf r},t)=\hat{e}_z\Omega({\bf r},t)$ and $\boldsymbol{\Omega}=\hat{e}_z\Omega$. Equation \eqref{eqn:nlr} has the form of a standard rotating conservation law for the probability density and does not depend on any additional terms that constitute loss or gain of probability density. From this we can conclude that the associated solutions are stationary, which we explore in the next section.}

\section{\label{sec:numerics}Numerical results}
\subsection{\label{sec:inr}Isotropic nonlinear rotation}
In this section we explore the numerical stationary-state solutions to the generalized Schr\"odinger equation, Eq.~\eqref{eqn:nlr} for different physical conditions. These solutions are calculated using a finite difference scheme, the details of which are provided in Appendix \ref{app:numerics}. Understanding the possible vortex configurations in superfluid systems remains an ongoing interest, see for example refs.~\cite{xie_2018,adhikari_2019,doran_2020} for recent studies.

In this subsection we begin by considering the possible ground state configurations of Eq.~\eqref{eqn:nlr} in the parameter space of the rigid body and nonlinear rotation strength $(\Omega,\mathcal{C})$ for fixed trap anisotropy $\omega_y/\omega_x=1.01$ and two-body mean-field strength $g_{\rm eff}N/\hbar\omega_x a_{x}^2=3\times 10^2$. \me{Note that in order to nucleate vortices in this system we require a finite rigid-body rotation strength ($\Omega>0$) since the vorticity would otherwise be zero at the edges of the system due to the presence of the trapping potential.} In Fig.~\ref{fig:vortnum} the number of vortices $N_v$ is presented in the $(\Omega,\mathcal{C})$ parameter space, allowing us to interpret the effect of these two parameters on both the morphology and topology of the superfluid state. In general, increasing $\Omega$ for fixed nonlinear rotation strength $\mathcal{C}$ increases the observed number of vortices. If instead we fix the rigid body rotation strength $\Omega$, we find that the number of vortices increases with the modulus of $\mathcal{C}$. The morphology of \me{the} superfluid is strongly affected by the choice of $\mathcal{C}$. For large positive $\mathcal{C}$ the vortices tend to localize, for example the ground state density $|\psi(x,y)|^2$ for $(\Omega/\omega_x=0.5,\mathcal{C}Nm/\hbar=20)$ and $(\Omega/\omega_x=0.65,\mathcal{C}Nm/\hbar=15)$ show tightly-packed arrangements (five and nine vortices respectively). On the other hand for large negative $\mathcal{C}$ the vortices tend to delocalize into ring structures, for example $(\Omega/\omega_x=0.55,\mathcal{C}Nm/\hbar=-25)$ and $(\Omega/\omega_x=0.6,\mathcal{C}Nm/\hbar=-35)$ each show a single ring of five and seven vortices respectively. As well as the ring arrangements, we also observe concentric ring configurations. For $(\Omega/\omega_x=0.65,\mathcal{C}Nm/\hbar=-10)$ a pair of rings formed from individual triangular vortex patterns is observed, while for $(\Omega/\omega_x=0.7,\mathcal{C}Nm/\hbar=-30)$ a pair of rings constructed from vortices occupying the vertices of two squares is found. It is also possible to find a ground state comprising a single ring of vortices surrounding a single vortex at the origin of the harmonic trap, as shown for $(\Omega/\omega_x=0.7,\mathcal{C}Nm/\hbar=-10)$. The other pair of panels depicted in Fig.~\ref{fig:vortnum} show the energy of the individual ground states computed from the definition
\begin{figure}[t]
\includegraphics[width=\columnwidth]{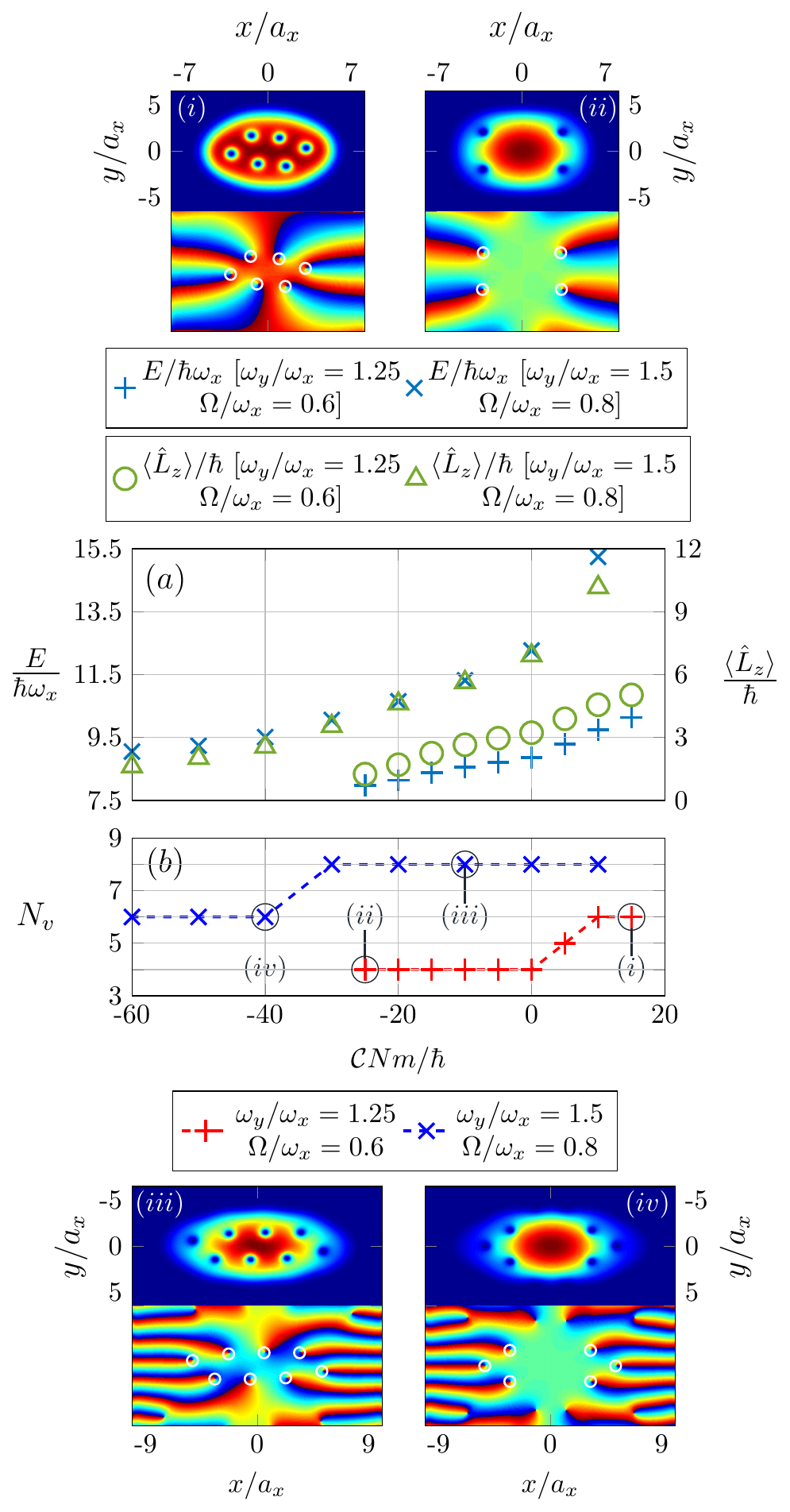}
\caption{\label{fig:ellirot}(color online) Nonlinear elliptical rotation ground states. Panel (a) shows the energy (blue pluses and crosses) calculated using Eq.~\eqref{eqn:energy} and angular momentum $\langle\hat{L}_z\rangle$ (green circles and triangles) computed for the trapping anisotropies $\omega_y/\omega_x=1.25,1.5$ as a function of the nonlinear rotation strength. Panel (b) shows the corresponding number of vortices $N_v$ as a function of $\mathcal{C}Nm/\hbar$. Four example ground state density and phases are also shown labeled $(i){-}(iv)$ in (b).}
\end{figure}
\begin{figure*}[t]
\centering
\includegraphics[scale=0.78]{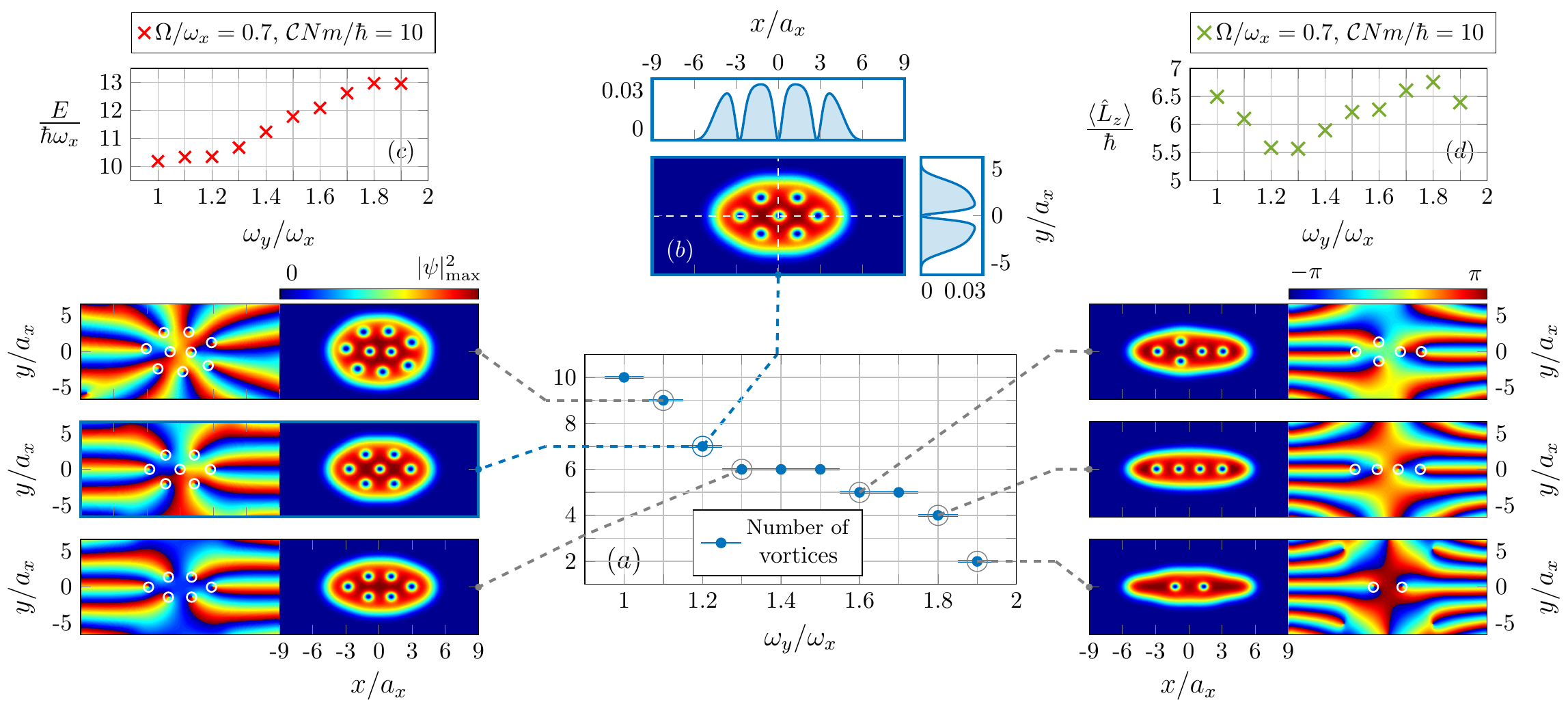}
\caption{\label{fig:anitrap}(color online) Deformed trap ground states. Panel (a) shows the number of vortices computed as a function of the trap anisotropy $\omega_y/\omega_x$, while the dashed arrows connect example ground states for different anisotropies. Example ground states show the density and phase, with vortices highlighted in the phase. Panel (b) shows the density for $\omega_y/\omega_x=1.2$ with accompanying cartesian cross-sections of the density. Panels (c) and (d) show the associated energy and angular momentum respectively. Throughout $\Omega/\omega_x=0.7$ and $\mathcal{C}Nm/\hbar=10$.}
\end{figure*}
\begin{equation}\label{eqn:energy}
E{=}\int d\boldsymbol{\rho}\bigg[\frac{\hbar^2}{2m}|\nabla\psi|^2{+}V|\psi|^2{+}\frac{g_{\rm eff}}{2}|\psi|^4{-}\boldsymbol{\Omega}\cdot\psi^{*}\hat{L}_z\psi\bigg],
\end{equation}  
which for a fixed value of $\Omega$ increases monotonically as a function of $\mathcal{C}$. Intuitively we can understand this behaviour since vortices at the edge of the cloud are localized in a low density region, and have a lower energy than those at larger values of $\mathcal{C}$ that may be closer to centre of the trap, possessing a corresponding larger ground state energy. The final panel (c) displays the mean value of the angular momentum, computed from $\langle\hat{L}_z\rangle=\int d\boldsymbol{\rho}\psi^{*}\hat{L}_z\psi$. Again the computed value of $\langle\hat{L}_z\rangle$ monotonically increases from negative to positive $\mathcal{C}$ for fixed $\Omega$, due to the effect of the background of the trapped cloud modulating $\langle\hat{L}_z\rangle$ with the position of the vortices. One can also understand the unusual vortex phenomenology present in this system by considering the (averaged) vorticity, defined as $\boldsymbol{\omega}_v=\nabla\times{\bf v}$  \cite{barenghi_2016}, where ${\bf v}=\Omega(\rho,t)\hat{e}_z\times\boldsymbol{\rho}$. Using these two definitions one can show that
\begin{equation}\label{eqn:wv}
\boldsymbol{\omega}_v=\hat{e}_z\bigg[2\Omega(\rho,t)+\rho\frac{\partial\Omega(\rho,t)}{\partial\rho}\bigg].
\end{equation}
The vorticity given by Eq.~\eqref{eqn:wv} varies in both time and space. This can be used to interpret the unusual vortex arrangements described in this work, here the second (novel) term appearing in Eq.~\eqref{eqn:wv} depends on the radial derivative of the atomic density (Eq.~\eqref{eqn:tfexp}), leading to a contribution to the vorticity that depends on the local radial curvature of the state.     

\subsection{\label{sec:etd}Elliptical trap deformation}
Next we explore the effect of varying the strength of the nonlinear rotation strength while fixing both the trap anisotropy and rigid body rotation strengths. \me{Although angular momentum is no longer conserved when $\omega_y\neq\omega_x$, stationary states of the generalized Gross-Pitaevskii model (Eq.~\eqref{eqn:nlr}) can still be obtained in the rotating frame in a similar spirit to Refs.~\cite{oktel_2004,mcendoo_2010} where stationary states were obtained for non-axisymmetric confined gases for finite rigid body rotation.} Figure \ref{fig:ellirot} explores the effect of computing the ground states of Eq.~\eqref{eqn:nlr} in two configurations -- the first with constant $\omega_y/\omega_x=1.25$ and $\Omega/\omega_x=0.6$ and the second with constant $\omega_y/\omega_x=1.5$ and $\Omega/\omega_x=0.8$. Note that the change in the rigid body strength here is to accommodate the fact that working at larger values of the trap anisotropy $\omega_y/\omega_x$ tends to reduce the observed number of vortices, hence $\Omega/\omega_x$ is slightly increased for the case of $\omega_y/\omega_x=1.5$. Then, Fig.~\ref{fig:ellirot}(a) shows the ground state energy $E$ (left axis) computed using Eq.~\eqref{eqn:energy} and corresponding angular momentum $\langle\hat{L}_z\rangle$ (right axis). The energy of the ground states, shown as blue pluses and crosses is found to be increasing from negative to positive values of $\mathcal{C}Nm/\hbar$, again as described in Sec.~\ref{sec:inr} this is due to the vortices wanting to delocalize at the edges of the harmonic trap for $\mathcal{C}\ll0$, resulting in a lower overall energy. Meanwhile for large (positive) $\mathcal{C}$ the vortices are localized closer to the centre of the trap, resulting in a (relatively) higher ground state energy. The angular momentum is shown as green circles and triangles and shows a similar trend to the energy, increasing from negative to positive $\mathcal{C}$. We also note that increasing the trap anisotropy $\omega_y/\omega_x$ has the effect of reducing the amount by which the energy and angular momentum vary. Panel (b) meanwhile shows the number of vortices $N_v$ for both configurations presented in (a). Interestingly we find that there exists a value of the nonlinear rotation strength $\mathcal{C}$ where the number of vortices is reduced but does not increase again, in contrast to the situation with cylindrical trap symmetry described in Sec.~\ref{sec:inr} previously. We attribute this behaviour to the breaking of conservation of angular momentum, since the density-dependent rotation in the trapped limit essentially depends on the shape of the harmonic confinement which becomes elliptical for $\omega_y\neq\omega_x$. Finally in Fig.~\ref{fig:ellirot} there are several example ground states, labelled $(i)$-$(iv)$ showing different vortical configurations for parameters shown in (b), with the top panels in each example showing the density $|\psi(x,y)|^2$ while the lower panels show the associated phase distribution $\vartheta(\boldsymbol{\rho})=\tan^{-1}\{\text{Im}(\psi)/\text{Re}(\psi)\}$. The individual vortices are highlighted in the phase with white circles.

We can also investigate the effect on the superfluid state by fixing both of the rotational parameters $\Omega$ and $\mathcal{C}$ and instead vary the ellipticity of the harmonic confinement through $\omega_y/\omega_x$. Numerical simulations of this situation are presented in Fig.~\ref{fig:anitrap} for $\Omega/\omega_x=0.7$ and $\mathcal{C}Nm/\hbar=10$. Panel (a) displays the number of vortices $N_v$ as a function of the trap anisotropy. As the trap geometry is gradually reduced from the quasi-two-dimensional ($\omega_y\sim\omega_x$) to quasi-one-dimensional ($\omega_y\gtrsim\omega_x$) limit the number of observed vortices reduces, consistent with the known behaviour of anisotropically confined phase defects \cite{mcendoo_2009}. Then, six individual ground state configurations are also presented for different values of the trap anisotropy $\omega_y/\omega_x$. We observe that close to the cylindrical $\omega_y\sim\omega_x$ limit the vortices adopt almost triangular configurations, while for $\omega_y/\omega_x=1.8,1.9$ the vortices instead adopt a one-dimensional alignment as well as being reduced in number. Panel (b) displays the ground state density $|\psi(x,y)|^2$ for $\omega_y/\omega_x=1.2$, as well as accompanying cartesian cross-sections $|\psi(x,y=0)|^2$ (above panel) and $|\psi(x=0,y)|^2$ (right panel) showing the cores of the vortices for this example. Panels (c) and (d) show the corresponding ground state energy $E$, Eq.~\eqref{eqn:energy} and angular momentum $\langle\hat{L}_z\rangle$ respectively. The ground state energy $E$ is observed to increase \me{monotonically with} $\omega_y/\omega_x$, again consistent with the known behaviour of these systems (viz. Eq.~\eqref{eqn:muz}). Finally panel (d) showing the angular momentum $\langle\hat{L}_z\rangle$ is observed to oscillate as the trap anisotropy is increased -- an unusual and unexpected result, since increasing $\omega_y/\omega_x$ typically causes a reduction of the angular momentum, since the number of vortices is reduced at larger trapping anisotropies. This is likely due to the non-trivial effect of nonlinear rotation, which directly couples the deformed atomic density and the angular momentum of the condensate.                    

\subsection{Vortex lattices and rings}
\begin{figure}[t]
\includegraphics[width=0.95\columnwidth]{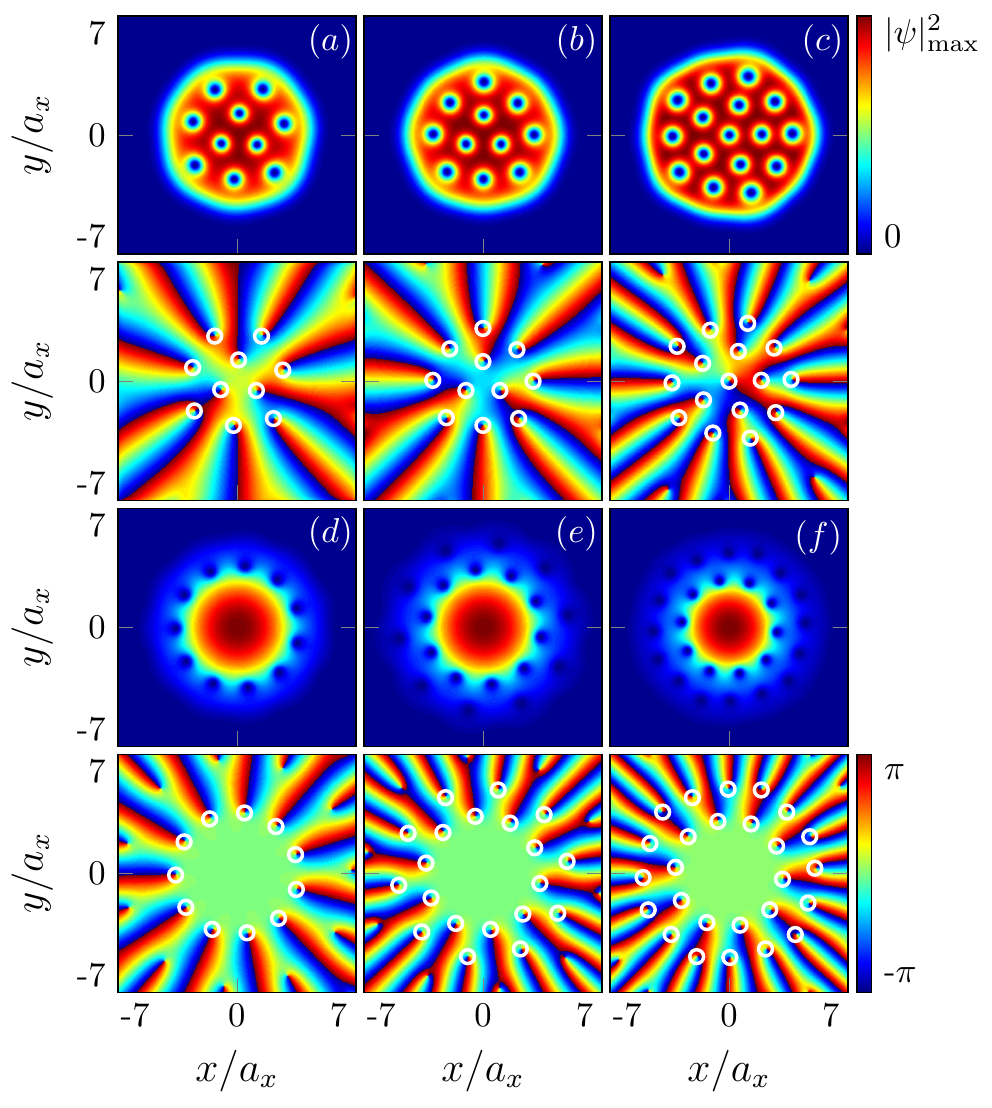}
\caption{\label{fig:lattice}(color online) Vortex lattices and rings. In panels (a)-(c) the atomic density $|\psi|^2$ is shown for fixed $\mathcal{C}Nm/\hbar=10$ and $\Omega/\omega_x=0.7,0.75,0.85$ rspectively, while the corresponding phase is shown in each panel below. Then in panels (d)-(f) the fixed nonlinear rotation strength is $\mathcal{C}Nm/\hbar=-60$ while $\Omega/\omega_x=0.75,0.8,0.85$. Again the phase distributions are shown below in each case. Throughout $\omega_y/\omega_x=1.01$. }
\end{figure}
In the final part of this section we examine the effect of the interplay of rigid body and nonlinear rotation when there is significant vorticity, for the almost cylindrical trap $\omega_y/\omega_x=1.01$. For the case of pure rigid body rotation ($\mathcal{C}=0$) it is well established that for large rotational driving strengths $\Omega\sim\omega_x$ dense triangular arrangements of vortices are observed. The presence of the nonlinear rotation in Eq.~\eqref{eqn:nlr} is expected to give rise to unusual many-vortex states, since we know that the presence of this term modifies the morphology of the vortex arrangements. Recent related work has also revealed the vortex patterns of bosonic systems with long-ranged statistical interactions \cite{correggi_2019}, a cousin of the model considered in this work. 

Figure \ref{fig:lattice} shows several example stationary states possessing significant vorticity. In panels (a)-(c) the atomic density is presented corresponding to the fixed nonlinear rotation strength $\mathcal{C}Nm/\hbar=10$ for increasing rigid body driving strengths $\Omega/\omega_x=0.7,0.75,0.85$. One can observe that the increasing number of vortices $N_v=10,11,16$ in these examples arrange into a triangular pattern, however it would seem that as $\Omega/\omega_x$  increases the density of the vortices also grows. We would speculate that this is an effect of the boundaries of the cloud, which as we move to higher rotational driving strengths become more deformed due to the nonlinear rotation and hence contribute more in this limit to the deformation of the observed vortex lattice. In each case, the corresponding phase distribution is shown in the panel directly below panels (a)-(c), with the vortices highlighted by white circles. Then panels (d)-(f) show stationary states obtained for fixed $\mathcal{C}Nm/\hbar=-60$, again with increasing rigid body rotation $\Omega/\omega_x=0.75,0.8,0.85$, with the respective number of vortices being $N_v=11,20,26$. Again for $\mathcal{C}<0$ we observe the spatial delocalization of the vortices, and absence of the triangular Abrikosov vortex lattice. Instead the vortices prefer to arrange into ring structures, as observed in (d). Interestingly ss $\Omega/\omega_x$ is increased, concentric (double) ring structures can also be obtained, as shown in (e) and (f). One can see that the shape of the background condensate is quite different for the examples presented in (a)-(c) with $\mathcal{C}>0$ and (d)-(f) with $\mathcal{C}<0$, due to the underlying generalized (rotating) density distribution $n(\rho)$, Eq.~\eqref{eqn:tfexp} possessing a distorted form compared to the standard Thomas-Fermi distribution. As before the corresponding phase distributions are presented below panels (d)-(f), and the vortices are highlighted with white circles.          

\section{\label{sec:summary}Summary}
 
In this work we have theoretically examined the few-to-many vortex states of a gas of bosons confined in an elliptical harmonic trap, subject to a density-dependent gauge potential that manifests as an effective density-angular-momentum coupling to the atomic condensate. By constructing analytical and semi-analytical solutions in the Thomas-Fermi limit, the stability of the nonlinear system was probed as a function of the anisotropy of the confinement, strengths of the nonlinear rotation and two-body interactions. This in turn revealed stable and unstable regions of the parameter space, the condensate being stable over the largest parameter values (nonlinear rotation strength) close to the cylindrical limit, while for more quasi-one-dimensional configurations the stable regions of the total parameter space were found to be more restricted.

Numerical simulations of the generalized Schr\"odinger equation revealed the vortices phenomenology, including the topology and morphology of the various ground states of the rotating system. Close to cylindrical confinement, different vortex patterns were presented exploring the interplay of the nonlinear rotation and the rigid body rotation. Vortex ring arrangements were observed and attributed to the effect of the unusual vorticity present in the system. Following this the \me{effect} of elliptical harmonic trapping was studied. Here, the trap aspect ration was fixed while the strength of the nonlinear rotation was varied, revealing a critical point where the vortices separate into opposing regions of the trapped cloud, an effect caused by the \me{spatially varying vorticity}. Following this the effect of fixing the nonlinear rotation strength and changing the trap ellipticity was investigated. It was found that similar to the case of rigid body rotation, increasing the trap anisotropy from a quasi-two to quasi-one-dimensional limit causes the total number of observed vortices to decrease. In the final part of this work we considered the effect of the density-angular-momentum coupling at larger rigid body rotation strengths in an almost cylindrical trap. It was found that for large positive nonlinear rotation strengths vortex lattices with increasingly densely arranged vortices are observed, while instead for large negative nonlinear rotation strength the vortices arrange into single ring and (multiple) concentric ring structures.

This work has explored the effect of elliptical density-angular-momentum coupling in quasi-two-dimensional atomic Bose-Einstein condensates, revealing the exotic phenomenology of this unusual physical system. \me{As well as being interesting from a fundamental physics perspective, exotic superfluid systems such as those considered in this work also provide an important potential resource for quantum technologies. One potential application in this realm is the emerging field of atomtronics which aims to harness cold atom systems to realize e.g. analogies of classical electronic circuits \cite{seaman_2007}. In this sense the density-dependent gauge theory studied in this work possesses a spatially varying vorticity which could be exploited to realize atomtronic circuits with radial symmetry without the need for complicated ring trap potentials \cite{eckel_2014}, due to the ability to realize  ring vortex arrangements in a standard harmonic trap.} In the future it would be interesting for example to understand experimentally motivated dynamical effects associated with dimensional crossovers, such as vortex-solitons and solitonic vortices, \me{as well as beyond mean-field effects such as density-dependent quantum droplets.} 
              
\section*{Acknowledgments}

M.E. is grateful to M. Nitta for discussions. This work is supported by Japan Society of Promotion of Science (JSPS) Grant-in-Aid for Scientific Research (KAKENHI Grants No. JP20K14376).

\appendix
\section{\label{app:numerics}Numerical simulations}
\begin{figure}[t]
\includegraphics[width=\columnwidth]{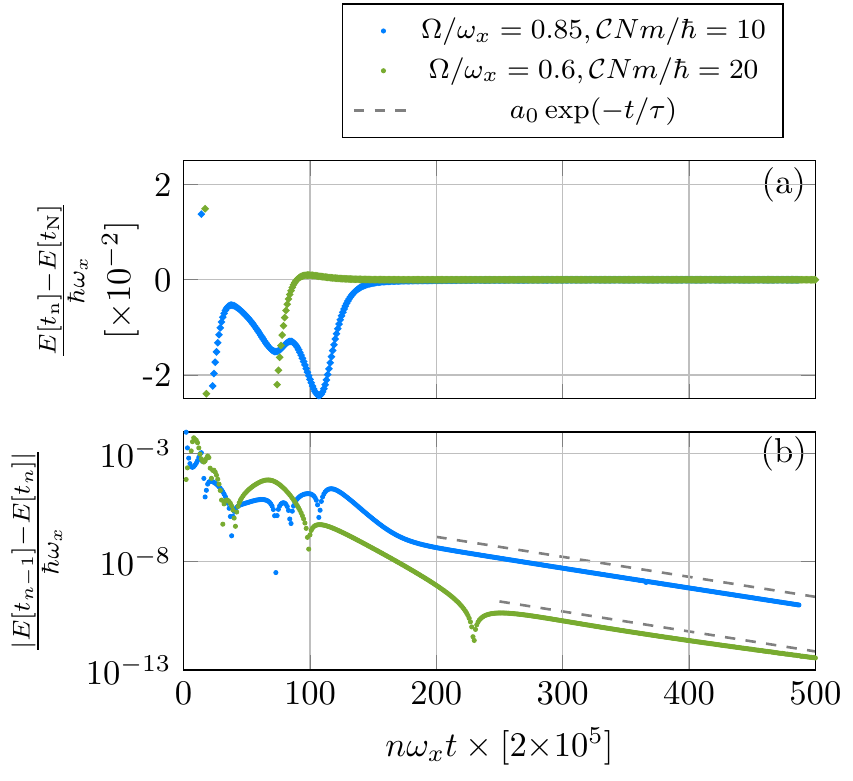}
\caption{\label{fig:itime}(color online) Imaginary time ground state calculations. Panel (a) shows example data for the difference of the sampled energy during imaginary time and the final energy, while (b) depicts the difference between the consecutively sampled energy. The two grey lines are a fit (see Eq.~\eqref{eqn:efit}) during the final equilibration of the vortex pattern.}
\end{figure}
In this appendix we outline the methodology used to compute the numerical solutions presented in Sec.~\ref{sec:numerics} of the paper. Essentially we propagate the generalized Schr\"odinger equation \eqref{eqn:nlr} in imaginary time using the discretization
\begin{align}\nonumber
&i\hbar\bigg[\frac{\psi^{n+1}_{j,k}-\psi^{n}_{j,k}}{\Delta t}\bigg]=\\\nonumber-&\frac{\hbar^2}{2m}\bigg[\frac{\psi^{n}_{j+1,k}{-}2\psi^{n}_{j,k}{+}\psi^{n}_{j-1,k}}{\Delta x^2}{+}\frac{\psi^{n}_{j,k+1}{-}2\psi^{n}_{j,k}{+}\psi^{n}_{j,k-1}}{\Delta y^2}\bigg]\\\nonumber&+V_{j,k}\psi^{n}_{j,k}+i\hbar\big(\Omega+\mathcal{C}|\psi^{n}_{j,k}|^2\big)\bigg[x_j\bigg\{\frac{\psi^{n}_{j,k+1}-\psi^{n}_{j,k-1}}{2\Delta y}\bigg\}\\&-y_k\bigg\{\frac{\psi^{n}_{j+1,k}-\psi^{n}_{j-1,k}}{2\Delta x}\bigg\}\bigg]+g_{\rm eff}|\psi^{n}_{j,k}|^2\psi^{n}_{j,k},
\end{align}
here the continuous wave function $\psi(x,y,t)$ becomes the discrete variable $\psi^{n}_{j,k}$. The spatial and temporal grid sizes are defined as $\Delta x,\Delta y$ and $\Delta t$ respectively, and numerical stability requires $\Delta t/\Delta\{x,y\}^2<\tfrac{1}{2}/\omega_{x}a_x^2$. In our simulations we used $\Delta x=\Delta y=0.05 a_x$ and $\Delta t=5\times10^{-5}/\omega_x$. 

Then, Fig.~\ref{fig:itime} shows data recorded from two example imaginary time runs for different physical parameters. Panel (a) shows $E[t_n]-E[t_N]$ i.e. the difference between the energy recorded after the $n^{\rm th}$ sample and the final sample $N$, where each sample is recorded after every $2\times10^{5}$ iterations. The lower panel (b) shows instead the difference between consecutive samples $E[t_{n-1}]-E[t_n]$ plotted semi-logarithmically. The initial condition $\psi^{0}_{j,k}$ for each run is taken as a pseudo-random matrix, which breaks any underlying symmetries and stops the simulation getting stuck in any metastable states. Note that for a finite (trapped) system vortices can only be nucleated if $\Omega\neq0$, since phase defects enter the cloud via the boundaries, so one must always work at finite $\Omega$ in order to generate vorticity with a density-angular-momentum coupling. Then, propagation in imaginary time leads to different dynamical regimes. For $n\omega_x t\lesssim100$, vortices enter the cloud from the edges of the system, and begin to arrange themselves inside the trap. After this for $100\lesssim n\omega_x t\lesssim200$ the vortices continue to arrange themselves closer to the equilibrium configuration, which occurs for long-times and is heralded by the exponential decay of the energy difference (dashed grey lines) which follows the exponential fit  
\begin{equation}\label{eqn:efit}
\frac{|E[t_{n-1}]-E[t_n]|}{\hbar\omega_x}=a_0\exp(-t/\tau), \hspace{0.5cm}n\gg1
\end{equation}
where $\tau=47/\omega_{x}$ and $a_0$ is a fitting parameter, which gives the converged ground state configuration. Individual simulations can be terminated after a certain tolerance of the energy difference is obtained, this is typically $\lesssim10^{-10}$ although configurations with more vortices present or where there are vortices closer to the edges of the harmonic trap require a smaller tolerance to produce the final high-fidelity ground state.

\end{document}